# Evaluation of new large area PMT with high quantum efficiency[*]

LEI Xiang-Cui (雷祥翠)[1,2,3;1)] HENG Yue-Kun (衡月昆)[1,2;2)] QIAN Sen (钱森)[1,2] XIA Jing-Kai (夏经铠)[1,2,3]
LIU Shu-Lin (刘术林)[1,2] WU Zhi (吴智)[1,2] YAN Bao-Jun (闫保军)[1,2] XU Mei-Hang (徐美杭)[1,2]
WANG Zheng (王铮)[1,2] LI Xiao-Nan (李小男)[1,2] RUAN Xiang-Dong (阮向东)[4] Wang Xiao-Zhuang (王小状)[1,5]
YANG Yu-Zhen (杨玉真)[1,6] WANG Wen-Wen (王文文)[1,6] FANG Can (方灿)[1,4] LUO Feng-Jiao (罗凤姣)[1,2,3]
LIANG Jing-Jing (梁静静)[1,4] YANG Lu-Ping (杨露萍)[1,7] YANG Biao (杨彪)[1,2]

1State Key Laboratory of Particle Detection and Electronics, Beijing 100049, China

2Institute of High Energy Physics, Chinese Academy of Sciences, Beijing 100049, China

3University of Chinese Academy of Sciences, Beijing 100049, China

4Guangxi University, Nanning 530004, China

5University of Science and Technology of China, Hefei 230026, China

6Nanjing University, Nanjing 200093, China

7Henan University, Kaifeng 475001, China

**Abstract:** The neutrino detector of the Jiangmen Underground Neutrino Observatory (JUNO) is designed to use 20 kilotons of liquid scintillator and approximately 16,000 20-inch photomultipliers (PMTs). One of the options is to use the 20-inch R12860 PMT with high quantum efficiency which has recently been developed by Hamamatsu Photonics. The performance of the newly developed PMT preproduction samples is evaluated. The results show that its quantum efficiency is 30% at 400 nm. Its Peak/Valley (P/V) ratio for the single photoelectron is 4.75 and the dark count rate is 27 kHz at the threshold of 3 mV while the gain is at $1 \times 10^7$. The transit time spread of a single photoelectron is 2.86 ns. Generally the performances of this new 20-inch PMT are improved over the old one of R3600.

**PACS:** 85.60.Ha    **DOI:** 10.1088/1674-1137/40/2/026002

## 1 Introduction

High performance large area PMTs are critical for large neutrino experiments to cover the large surface areas of the detectors in order to collect photons produced by neutrino interactions with detector media. Jiangmen Underground Neutrino Observatory (JUNO), which is under construction in south China, will use 20,000 tons of liquid scintillator which requires approximately 16,000 20-inch PMTs. To achieve the main goal of JUNO, to determine the mass hierarchy of neutrino generations, precision measurement of the neutrino energy spectrum with extremely high energy resolution of $3\%/\sqrt{E(MeV)}$ or better is required [1]. In order to collect sufficient neutrino events, the target mass of the central detector is as large as 20 kilotons. The range of the energies of scintillating photons from inverse beta decay of neutrino interactions is from approximately 1 to 10 MeV and must be collected with an efficiency as high as possible and their energies measured as precisely as possible in order to resolve the subtle changes in the neutrino energy spectrum caused by different mass hierarchy assumptions. Hamamatsu Photonics has developed a new type of large area dynode PMT with high quantum efficiency, the R12860, which is being considered as a candidate for JUNO. Here we report the quantum efficiencies of three preproduction samples of the R12860 PMTs, labeled as 1#, 2# and 3#, as well as single photoelectron response and anode dark count rate of PMT 1#.

## 2 Performance studies

The R12860 has a box-type first dynode followed by nine linear focused dynodes with a new type of high efficiency bialkali photocathode [2]. In this section, the measurement principle and setup are described and results for critical performance parameters of the R12860 PMT samples, including quantum efficiency, single photoelectron (SPE) response, gain, P/V ratio, resolution of the charge spectrum, dark pulse rate and timing properties are reported.

[*] This work is supported by the "Strategic Priority Research Program" of the Chinese Academy of Sciences (Grant No.X-DA10010200), the key deployment project of the Chinese Academy of Sciences and in part by the CAS Center for Excellence in Particle Physics (CCEPP).

1) E-mail: leixc@ihep.ac.cn

2) E-mail: hengyk@ihep.ac.cn

## 2.1 Quantum efficiency

The quantum efficiency (QE) is an important parameter for PMTs. Incident photons on the photocathode knock out electrons through the photoelectric effect, and the electrons emitted are called photoelectrons. The quantum efficiency is defined as the fraction of incident photons that are converted into photoelectrons. The other factor that influences the photon detection efficiency is the photoelectron collection efficiency, which is the fraction of photoelectrons leaving the photocathode, reaching the first dynode and being amplified by the dynodes. The higher the quantum efficiency and the photoelectron collection efficiency, the more photons can be detected and the better the energy resolution of the neutrino detector.

The experimental setup for QE measurement is shown in Fig. 1. The light source is a tungsten lamp. The light from the tungsten lamp first goes through a monochromator for wavelength selection. The focusing electrodes and the first dynode of the R12860 are connected together, and the photocurrent is measured. The reference photo diode (PD) quantum efficiency was calibrated in advance in the experiment. The picoammeter is controlled by Computer 2 and the monochromator is controlled by Computer 1. The voltages for the PD and PMT are supplied by the picoammeter. From the ratio of the currents from the PMT and PD, the QE of the PMT can be obtained. The QE of the PMT can be expressed as:

$$\eta_{PMT} = \eta_{PD} \frac{I_{PMT}}{I_{PD}}, \qquad (1)$$

where $\eta_{PMT}$ is the QE of the PMT; $\eta_{PD}$ is the QE of the reference PD as provided by the PD manufacturer; $I_{PMT}$ is the output current of the PMT and $I_{PD}$ is the output current of the PD.

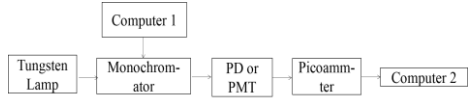

Fig. 1. Experimental setup for QE measurements

By changing the wavelength of the light with the monochromator and recording the current from the focusing electrodes and first dynode of the PMT, the relation of QE to wavelength is obtained. Fig. 2 shows the curve of QE versus wavelength for the PMT photocathode. The measured QE is higher than 25% in the wavelength range from 350 nm to 450 nm and at 400 nm, the QE reaches a maximum of about 30%. Differences among the three PMT samples are small.

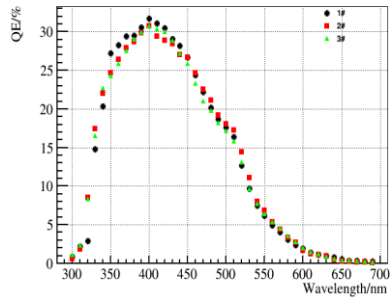

Fig. 2. QE versus wavelength for the three R12860 sample tubes

## 2.2 SPE response

Single photoelectron measurement capability is a critical requirement for the 20-inch PMTs in JUNO. The neutrino energy spectrum is measured by counting the total number of photoelectrons in each neutrino interaction. The PMT gain must be high enough to detect a single photoelectron. The PMTs must also be able to distinguish single photoelectron signals from the noise, which requires the single photoelectron peak to be as narrow as possible and have a large P/V ratio. The result in Section 2.2 is based on the PMT 1#, as well as Section 2.3.

**2.2.1 Gain, Resolution, Peak/Valley ratio**

The gain, SPE resolution, P/V ratio and detection efficiency can be obtained from the single photoelectron spectrum. Fig. 3 shows the SPE experimental setup. Ch1 of the pulse generator supplies the working voltage for the laser diode (LD) and Ch2 supplies the gate for QDC V965 after discriminated and delayed. Ch1 and Ch2 are synchronized by a pulse generator. The light from the LD illuminates the photocathode of the PMT and then the signal output from the anode of the PMT is sent to QDC V965. The data is recorded by the data acquisition system in the computer.

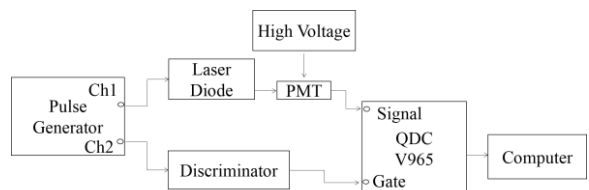

Fig. 3. Setup of SPE experiment

The fitting function for the SPE spectrum (Fig. 4) is described in Equation (2). The photoelectrons emitted from the photocathode and the first dynode can be described by the Poisson distribution [3] [4]. The output signals of the PMT can be expressed in the following form [5]:

$$S_{real}(x) \approx S_{ped} + S_{noise} + S_1 + S_m. \quad (2)$$

In Equation (2), $S_{ped}$ is the background caused by leakage current and other factors, which can be expressed as:

$$S_{ped} = \frac{1-w}{\sigma_0\sqrt{2\pi}} e^{-\frac{(x-Q_0)^2}{2\sigma_0^2}} e^{-\mu}. \quad (3)$$

$S_{noise}$ is the background caused by the dynode thermoemission and other factors and can be expressed as:

$$S_{noise} = w\theta(x - Q_0) \times \alpha e^{-\alpha(x-Q_0)} e^{-\mu}. \quad (4)$$

$S_1$ is the SPE signal and can be expressed as:

$$S_1 = \frac{\mu e^{-\mu}}{1!} \times \frac{1}{\sigma_1\sqrt{2\pi}} \times e^{-\frac{(x-Q_0-Q_{sh}-Q_1)^2}{2\sigma_1^2}}. \quad (5)$$

$S_m$ is the multi-photoelectron signal, which can be expressed as:

$$S_m = \sum_{n=2}^{\infty} \frac{\mu^n e^{-\mu}}{n!} \frac{1}{\sigma_1\sqrt{2\pi n}} e^{-\frac{(x-Q_0-nQ_1-Q_{sh})^2}{2n\sigma_1^2}}. \quad (6)$$

$Q_0$ is the mean value of the pedestal; w is the probability of the PMT background; $\alpha$ is the coefficient of the exponentially decreasing background of the PMT; $\theta(x)$ is a step function, $\theta(x - Q_0) = \begin{cases} 0, & x < Q_0 \\ 1, & x \geq Q_0 \end{cases}$; $\sigma_0$ is the standard deviation of the pedestal; $\sigma_1$ is the standard deviation of SPE; $Q_1$ is the PMT gain coefficient multiplied by the charge of one electron; and $Q_{sh}$ is the effective spectrum shift due to background.

The voltage supplied for the LD is adjusted until the percentage of the signal is about 10% in the experiment [6]. In this case, the percentage of pedestal is about 90%. According to the Poisson law $P(n) = \frac{\mu^n e^{-\mu}}{n!}$, suppose $e^{-\mu} = 0.9$, $\mu = 0.1054$; then, $P(1) = \mu e^{-\mu} = 0.09486$. The value of the ratio of SPE (P(1)) divided by the ratio of multi-photoelectron (P(2)) is $\frac{P(1)}{P(2)} = 18.46$. In this case, the signal can be referred to SPE signal.

The gain can be obtained from the charge spectrum and it can be expressed as the following:

$$G = \frac{Q_1}{e}. \quad (7)$$

G is the gain of the PMT; $Q_1$ is the charge of signal after subtracting the pedestal; and $e$ is the charge of one electron.

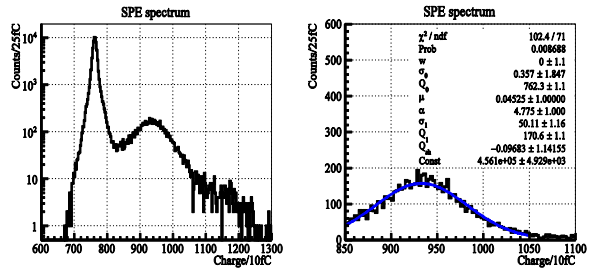

Fig. 4. The SPE spectrum

In the SPE mode, the $S_m \approx 0$. $Q_{sh}$ shown in Fig. 4 is small compared with $Q_1$, which means noise in this experiment is small. The ratio of SPE in the signals is 97.8%, according to the fitting results in Fig. 4. The gain is $\frac{170.6 \times 10 \times 10^{-15} C}{e} = 1.066 \times 10^7$, the resolution is $\frac{50.11}{170.6} = 29.37\%$ derived from the fitting curve, and the P/V ratio is 4.75.

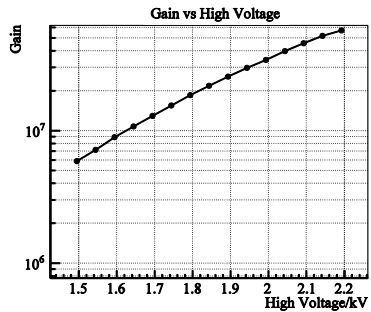

Fig. 5. HV versus gain curve

The HV versus gain curve is shown in Fig. 5. When the HV is above 1630 V, the gain is higher than $10^7$. The HV versus gain curve is consistent with an exponential relationship over the HV range from 1500 V to 2200 V.

To study the uniformity of the PMT, the LD is used to illuminate ten points of the photocathode separately, at the same longitude but different latitudes. The result is shown in Fig. 6. The relative gain ranges from 0.8 to 1 and the relative detectivity ranges from 0.15 to 1. When the latitude is about 36.26°, the relative detectivity is low. The reason is that the quality of the photocathode evaporation is poor when the latitude is lower than 43.1°.

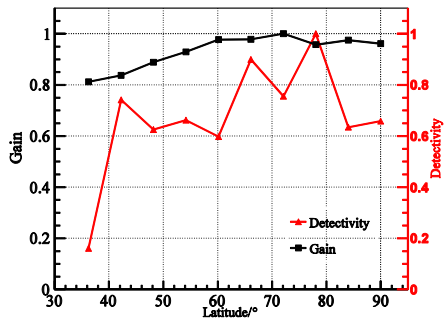

Fig. 6. The uniformity of different latitudes

The position of latitude on the photocathode is shown in Fig. 7. The outer arc (solid curve) indicates the surface of the input window, and the inner arc (dotted curve) is the photocathode. The latitude of 90° is the vertex of the photocathode, and the latitude of 43.1° is the edge of the effective area of the photocathode.

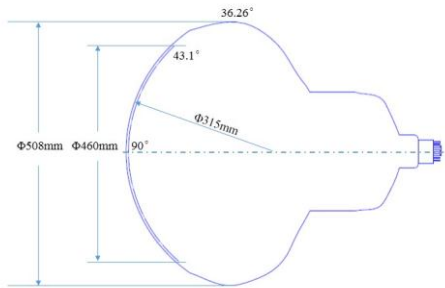

Fig. 7. The position of latitude on the photocathode

**2.2.2 Timing characteristics**

The neutrino event vertex reconstruction needs information obtained from the PMT single photoelectron timing measurements. The timing jitter influences the precision of reconstruction position of the vertex. Transit time spread (TTS) of SPE signals is the fluctuation of the transit time of single photoelectron signals measured by PMTs [7]. The setup for SPE TTS measurement is shown in Fig. 8.

Ch1 of the pulse generator supplies the working voltage for the laser diode which supplies the light for PMT. The output of the PMT is connected to the signal channel in the oscilloscope, which has a sampling rate of 10 GHz, and the trigger pulse in the oscilloscope is supplied by Ch2. The data recorded in the oscilloscope is sent to the computer.

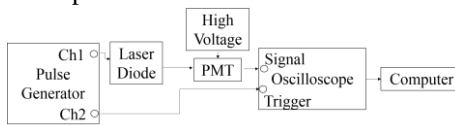

Fig. 8. Setup of TTS experiment

The driving voltage of the LD is adjusted to make the PMT work in SPE mode. A typical waveform of the PMT working in this mode is shown in Fig. 9, which is acquired by the oscilloscope. The amplitude shown in Fig. 9 is 6.3 mV. The rise time (6.10 ns) and the fall time (12.71 ns) can be obtained by the analysis of 200,000 digitized waveforms.

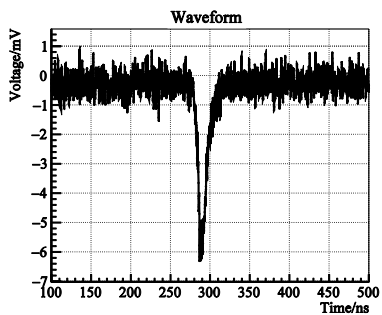

Fig. 9. The SPE waveform

The FWHM in the time spectrum is used to express the TTS. When tested, the gain is kept at $10^7$ and the triggering threshold for the signal is the amplitude of 0.25 photoelectron (pe). The time over threshold ($t_{measured}$) of the leading edge of the signal is calculated. $t_{measured}$ has a jitter for different amplitudes of the signal. A time-to-amplitude (T-A) correction is done to eliminate the time walk effect. The real time ($t_{real}$) after T-A correction can be expressed as below [8]:

$$t_{real} = t_{measured} - f(A). \quad (8)$$

$f(A)$ is the T-A correction function.

$$f(A) = p0 + \frac{p1}{\sqrt{A}} \quad (9)$$

The T-A correction curve is shown in Fig. 10.

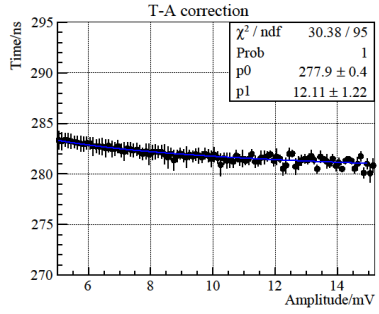

Fig. 10. T-A correction

The result after T-A correction is shown in Fig. 11. The TTS (FWHM) before T-A correction is 3.40 ns and after T-A correction, it is 2.86 ns. In fact, TTS is also influenced by the position of the cathode which the photons hit. The TTS value is more than 2.86 ns when the light illuminates the whole cathode.

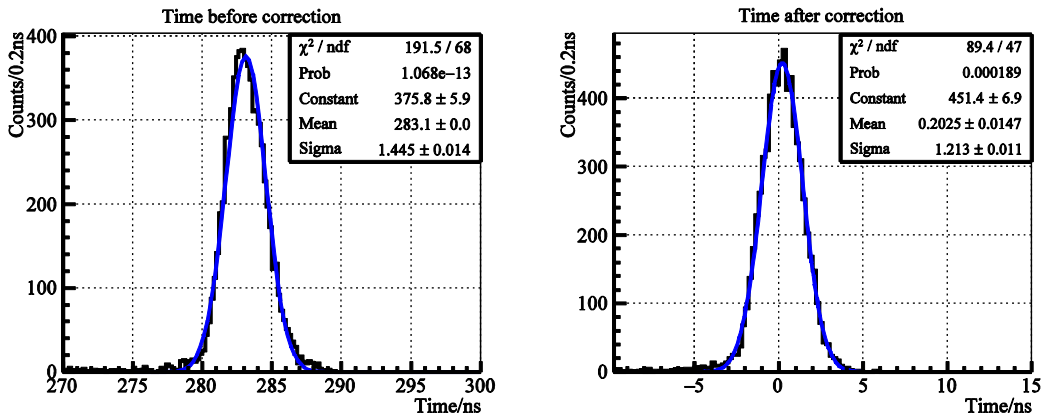

Fig. 11. The TTS of PMT 1#, left: before T-A correction, right: after T-A correction

## 2.3 Anode dark count rate

The SPE energy resolution and the properties of the neutrino signals require that the background of the PMT should be low. The anode dark count rate is used to evaluate the background level. There are several causes for dark counts, including thermionic emission from the photocathode and dynode, leakage current in the PMT, gamma rays from the environment, space electromagnetic interference, and the temperature according to Richardson's law [7][9]. The experiment was performed at room temperature and the gain kept at $10^7$. The triggering threshold ranged from 2 mV to 5 mV, as shown in Fig. 12. The data was saved every two minutes and the result in Fig. 12 shows the time-dependence curve. The count rate is high at the beginning as there is still light left and the working state of the PMT is not stable. The rate at this stage is not purely the anode dark rate. The count rate decreases as time goes on, taking several hours for the dark rate to settle down. The result in Fig. 12 shows that the anode dark count rate is about 27 kHz when the threshold is 3 mV.

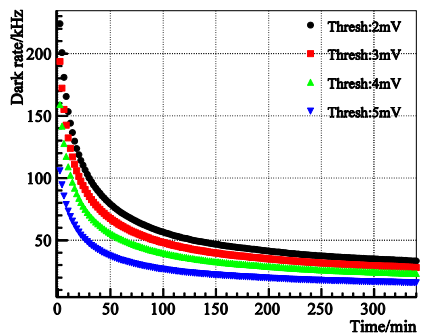

Fig. 12. Anode dark count rate for PMT 1#

## 3 Summary

The performance of the 20-inch Hamamatsu R12860 PMT preproduction samples has been evaluated. The quantum efficiency is approximately 30% at 400 nm, which is higher than traditional PMTs with bialkali photocathodes. The relation of gain versus high voltage, P/V value, resolution and TTS of the SPE were measured. The P/V ratio is 4.75 when the gain is $10^7$. T-A correction was determined in the TTS measurement, and the TTS decreased from 3.40 ns to 2.86 ns after the T-A correction. The anode dark count rate is less than 30 kHz when the threshold of the SPE trigger is 3 mV. These results provide important information for JUNO and other similar future neutrino experiments in selecting their photon detectors.